\documentclass[twocolumn,aps,pra,floatfix,groupedaddress,longbibliography]{revtex4-1}

\usepackage{xcolor}
\usepackage[utf8]{inputenc} 
\usepackage{bm}
\usepackage{amsmath,amsthm,amssymb}
\usepackage{graphicx} 
\usepackage{pgfplots}
\usepackage{tikz,tkz-tab}{{}}
\usepackage[
  colorlinks=true,
  urlcolor=blue,
  linkcolor=blue,
  citecolor=blue
]{hyperref}
\usepackage[normalem]{ulem}
\usepackage{stackrel}

\def\urlprefix{}
\def\url#1{}

\newcommand{\ee}{\mathrm{e}}
\newcommand{\ii}{\mathrm{i}}
\newcommand{\ini}{\mathrm{ini}}

\newcommand{\mean}[1]{\langle#1\rangle}

\long\def\soutjd#1{}

\newcommand{\bs}{\boldsymbol}

\begin{document}

\title{Coherent seeding of the dynamics of a spinor Bose-Einstein condensate:\\ from quantum to classical behavior}

\author{Bertrand Evrard, An Qu, Jean Dalibard and Fabrice Gerbier}

\affiliation{Laboratoire Kastler Brossel, Coll{\`e}ge de France, CNRS, ENS-PSL Research University, Sorbonne Universit{\'e}, 11 Place Marcelin Berthelot, 75005 Paris, France}

\date{\today}
\pacs{}

\begin{abstract}
We present experiments revealing the competing effect of quantum fluctuations and of a coherent seed in the dynamics of a spin-1 Bose-Einstein condensate, and discuss the relevance of a mean-field description of our system. We first explore a near-equilibrium situation, where the mean-field equations can be linearized around a fixed point corresponding to all atoms in the same Zeeman state $m=0$. Preparing the system at this classical fixed point, we observe a reversible dynamics triggered by quantum fluctuations, which cannot be understood within a classical framework. We demonstrate that the classical description becomes accurate provided a coherent seed of a few atoms only is present in the other Zeeman states $m= \pm 1$. In a second regime characterized by a strong non-linearity of the mean-field equations, we observe a collapse dynamics driven by quantum fluctuations. This behavior cannot be accounted for by a classical description and persists for a large range of initial states. We show that all our experimental results can be explained with a semi-classical description (truncated Wigner approximation), using stochastic classical variables to model the quantum noise.

\end{abstract}

\maketitle

\section{Introduction} 

The mean-field approximation is an essential tool of  many-body physics. In this approach, the interaction of a single body with the rest of the system is treated in an averaged way, neglecting fluctuations around the mean and erasing any spatial correlations. The original many-body problem is then reduced to a much simpler one-body problem, a tremendous simplification enabling a basic analysis of the problem at hand. The accuracy of the averaging improves with the number of particles in direct interaction.  Consequently, the mean-field treatment is well suited for highly connected systems, while important deviations are common for systems with short range interactions in reduced dimensions. 

When applied to bosonic quantum systems, a mean-field approach often entails another important approximation where intrinsic quantum fluctuations (and the correlations they induce) are neglected. Since quantum fluctuations are reflected in the non-commutativity of observables, field operators in the second-quantization formalism are replaced by commuting $c$-numbers. A possible improvement consists in replacing the field operators by classical stochastic fields\,\cite{gardiner2004,polkovnikov2010,steel1998,sinatra2002,mathew2017}, with a statistics properly chosen to be as close as possible to the original quantum problem. Such a semi-classical approach allows to account quantitatively for quantum fluctuations, while keeping the inherent simplicity of the mean-field equations.


In this Letter, we study the role of quantum fluctuations and the emergence of mean-field behavior in a quantum spinor Bose-Einstein condensate \cite{kawaguchi2012}. The atoms are condensed in the same spatial mode and interact all-to-all. The mean-field approach is thus well appropriate to study the dynamics in the spin sector, and has indeed been successfully used to describe several situations, either at\,\cite{zhang2003a,jacob2012a} or out-of\, \cite{zhang2005a,chang2005a,kronjaeger2005a,kronjaeger2006a,black2007a,liu2009} equilibrium.
More recently, several experiments addressed the dynamics of a condensate prepared in an unstable configuration, achieving a high sensitivity to both classical and quantum fluctuations\,\cite{klempt2010,bookjans2011,lucke2011,hamley2012,lucke2014,linnemann2016,linnemann2017,kunkel2018,fadel2018,lange2018,Tian2020,Yang2019,qu2020,mias2008,cui2008a,wrubel2018,evrard2020}. 

Here, our goal is twofold. First, we reveal the effect of quantum fluctuations in two different dynamical regimes, corresponding to persistent oscillations or relaxation to a stationary state \cite{evrard2020}. Second, we address the relevance of a classical field description by comparing our experimental results systematically with three theoretical approaches. In the fully classical picture (C), we derive mean-field equations of motion and solve them for well-defined initial conditions, possibly including a coherent seed. In the semi-classical picture (SC), we keep the same mean-field equations of motion but for fluctuating initial conditions, with a probability distribution designed to model the quantum noise of the initial state. Finally, we perform a fully quantum treatment (Q), consisting in a numerical resolution of the many-body Schrödinger equation.

\section{Spinor Bose-Einstein condensates}\label{sec:spinorbec}

We work with Bose-Einstein condensates of $N$ spin-1 sodium atoms in a tight optical trap. Due to the strong confinement, all atoms share the same spatial wave function $\psi({\bs r})$ \cite{yi2002}, such that the spin is the only relevant degree of freedom. In this regime, the Hamiltonian describing the spin-spin interaction is (up to an additive constant)\,\cite{ohmi1998,ho1998,yi2002,kawaguchi2012}
\begin{align}
\hat{H}_{\rm int}=\frac{U_s}{2N}\sum_{i,j=1}^N\hat{\bs s}_i\cdot\hat{\bs s}_j=\frac{U_s}{2N}\hat{\bs S}^2\,.\label{eq.Hamiltonian_int}
\end{align}
Here $\hat{\bs s}_i$ denotes the spin of atom $i$, $\hat{\bs S}=\sum_i \hat{\bs s}_i$ the total spin, and $U_s$ the spin-spin interaction energy. In the single-mode limit, the spin-spin interaction is given by $U_s=(4\pi\hbar^2a_s N/M)\int d^3{\bs r}\,\vert\psi({\bs r})\vert^4$, where $a_s$ is a spin-dependent scattering length, $M$ is the mass of a sodium atom, and the spin-independent spatial mode $\psi$ is the lowest energy solution of the time-independent Gross-Pitaevskii equation\,\cite{dalfovo1999}. Note that technical fluctuations of the atom number $N$ translate into fluctuations of $U_s$ (other factors, such as fluctuations of the trap geometry, can also contribute to the latter). As will be discussed in more detail in Section\,\ref{sec: relaxation}, these technical fluctuations add to the intrinsic relaxation due to quantum fluctuations and thereby play a significant role in the interpretation of the experiments.

We use a magnetic field ${\bs B}$ aligned along the $z$ axis to shift the energies of the individual Zeeman states $|m\rangle$, the eigenstates of $\hat{s}_z$ with eigenvalues $m=0,\pm1$. Up to second order in $B$, the Zeeman Hamiltonian is $
\hat H_Z=\sum_{i=1}^N p\hat s_{zi}+ q\hat s_{zi}^2\,$,
where $p\propto B$ and $q\propto B^2$ are the linear and quadratic Zeeman shifts, respectively. 
Noticing that $[\hat S_z,\hat H_{\rm int}]=0\,$, the first term in $\hat H_Z$ is a constant of motion that can be removed by a unitary transformation. The total Hamiltonian thus reads\,\cite{kawaguchi2012}
\begin{align}
\hat{H}=\hat{H}_{\rm int}+\hat H_Z=\frac{U_s}{2N}\hat{\bs S}^2+ q\left(\hat{N}_{+1}+\hat N_{-1}\right)\,,\label{eq.Hamiltonian}
\end{align}
where $\hat N_m$ is the number of atoms in $|m\rangle$. 

Under a mean-field approximation, the annihilation operators $\hat a_m$ are replaced by the $c$-numbers $\sqrt{N}\zeta_m=\sqrt{N_m}\exp(i\phi_m)$. By convention we set $\phi_0=0$, and we focus on the situation $S_z=0$. We define the mean number of $(+1,-1)$ pairs $N_{\rm p}=({N}_{+1}+N_{-1})/2$, and take its normalized value $n_{\rm p}=N_{\rm p}/N$ and the conjugate phase $\theta=\phi_{+1}+\phi_{-1}$ as dynamical variables. In terms of these variables, the mean-field equations of motion are \cite{zhang2005a}
\begin{align}
\hbar\dot n_{\rm p}&=-2U_sn_{\rm p}(1-2n_{\rm p})\sin\theta\,,\label{eq: dnp/dt}\\
\hbar\dot \theta&=-2q+2U_s(4n_{\rm p}-1)(1+\cos\theta)\,.\label{eq: dtheta/dt}
\end{align}
At $t=0$, the BEC is prepared in a generalized coherent spin state $|\psi_\ini\rangle=\left(\sum_m\zeta_{\ini,m}|m\rangle\right)^{\otimes N}$, with
\begin{align}
	{\bs \zeta}_\ini=\begin{pmatrix}
	\sqrt{n_{\rm seed}}\,\ee^{i\frac{\theta_\ini+\eta_\ini}{2}}\\
	\sqrt{1-2n_{\rm seed}}\\
	\sqrt{n_{\rm seed}}\,\ee^{i\frac{\theta_\ini-\eta_\ini}{2}}\\
	\end{pmatrix}\,,\label{eq: coherent state}
\end{align}
where $n_{\rm seed}=N_{\rm seed}/N$ and $N_{\rm seed}$ is the number of atoms in the $m=\pm 1$ states.  The Larmor phase $\eta=\phi_{+1}-\phi_{-1}$ evolves as $\eta(t)=\eta_\ini-2pt/\hbar$ and does not play any important role in the following. We focus on  the behavior of $N_{\rm p}(t)$ as a function of time.


We notice that the state with all atoms in $m=0$ (\textit{i.e.} $N_{\rm seed}=0$ and hence $n_{\rm p}=0$) is stationary according to Eq.\,(\ref{eq: dnp/dt},\ref{eq: dtheta/dt}). However, this state is not an eigenstate of $\hat H_{\rm int}$ and thus not a stationary state of the quantum equation of motion. In the absence of a seed, we identified in Ref.\,\cite{evrard2020} two different regimes for the ensuing non-classical dynamics:
\begin{itemize}
\item For $U_s/N\ll q$, the dynamics is reversible: The number of pairs $N_{\rm p}(t)$ oscillates with a small amplitude. 
\item For $q\ll U_s/N$\,, the dynamics is strongly damped and $N_{\rm p}(t)$ relaxes to a stationary value. 
\end{itemize}
Here, we revisit these experiments to investigate the effect of a coherent seeding of the $m=\pm 1$ modes.

\section{Reversible dynamics}\label{sec: reversible}
\paragraph{Theoretical predictions}

We focus first on the situation where $U_s/N\ll q\ll U_s$ and $n_{\rm seed}\ll1$.
In this case, the reduced number of pairs $n_{\rm p}$ remains small at all times. Linearizing the mean-field Eqs.\,(\ref{eq: dnp/dt},\ref{eq: dtheta/dt}), we obtain\,\cite{SM}
\begin{align}
N_{\rm p}^{\rm (C)}(t)\approx \frac{2U_s}{q}\sin^2(\omega t)\cos^2\left(\frac{\theta_\ini}{2}\right)N_{\rm seed}\,,\label{eq: N_p_mf rev}
\end{align}
where $\omega\approx \sqrt{2qU_s}$.  Note that the oscillation frequency $\omega$ is independent on the initial conditions $\theta_{\ini}$ and $N_{\rm seed}$. In Sec.\,\ref{sec: relaxation}, we investigate a regime, where the frequency of the classical solution increases with $N_{\rm seed}$, with dramatic consequences on the semi-classical dynamics.

To improve the prediction (\ref{eq: N_p_mf rev}) and account for quantum fluctuations, we use a semi-classical approach, the truncated Wigner approximation \cite{polkovnikov2010,steel1998,sinatra2002,mathew2017,wrubel2018}. The probability amplitudes $\zeta_{\mathrm{ini},m}$ are treated as complex random variables which sample the initial Wigner distribution of the initial state at $t=0$. The amplitudes are then propagated according to the mean-field equations of motion. Averaging the mean-field predictions over the fluctuations of $\bm{\zeta}_{\rm ini}$, we find\,\cite{SM,wrubel2018}
\begin{align}
N_{\rm p}^{\rm (SC)}(t)\approx \frac{U_s}{2q}\sin^2(\omega t)\left[4\cos^2\left(\frac{\theta_\ini}{2}\right)N_{\rm seed}+1\right]\,.\label{eq: N_p_TWA rev}
\end{align}
In analogy with quantum optics, the term $\propto N_{\rm seed}$ in Eqs.\,(\ref{eq: N_p_mf rev},\ref{eq: N_p_TWA rev}) describes ``stimulated emission" from the mode $m=0$ to the modes $m=\pm1$, while the additional term in Eq.\,(\ref{eq: N_p_TWA rev}) can be interpreted as ``spontaneous emission". We have verified numerically that the SC results are in good agreement with a fully quantum treatment. Moreover, comparing equations (\ref{eq: N_p_mf rev}) and (\ref{eq: N_p_TWA rev}), we notice that unless the initial phase is chosen such that $\theta_\ini\approx\pi$, a large seed $N_{\rm seed}\gg1$ makes the C and SC treatments almost identical, irrespective of the precise value of $N$. In fact, seeding with a few atoms $N_{\rm seed}\approx 2-3$ and with $\theta=0$ is sufficient to reach a 90\,\% agreement between the two approaches.

\paragraph{Experimental sequence} 

We prepare a BEC in the state $m=0$ using evaporative cooling in a crossed laser trap with a large magnetic field $B=1\,$G ($q\gg U_s$). After evaporation, the BEC contains $N\approx 2000$ atoms in the state $m=0$, with $N_{\rm p}\approx100$ residual thermal atoms in $m=\pm1$. We then turn on a strong magnetic field gradient to pull the $m=\pm1$ atoms out of the trap. After this purification step, we measure $N_{\rm p}\ll 1$\,\cite{qu2020}. 

We add a coherent seed using a combination of magnetic field ramps and resonant radio frequency (rf) pulses.  In a first step, a rf pulse is used to prepare the atoms in a coherent superposition with a probability $n_{\rm seed}$ to be in a given $m=\pm1$ state. In a second step, the BEC is held in a large magnetic field, such that $q\gg U_s$ and $\theta_\ini$ can be tuned keeping $n_{\rm p}=n_{\rm seed}$ (see Supplementary Material\,\citep{SM} for more details). In this way, we are able to prepare any coherent spin state given by Eq.\,(\ref{eq: coherent state}), up to the phase $\eta_\ini$ which is irrelevant for the experiments described here. The main imperfection in the preparation originates from the fluctuations of the total atom number $\delta N\approx 0.1\,N$, which induce $\approx10\%$ relative fluctuations on $N_{\rm seed}$. 
%
 The magnetic field is then quenched to the desired value, and we let the system evolve for a time $t$ before measuring the population of each Zeeman state using a combination of Stern-Gerlach separation and fluorescence imaging with a detection sensitivity around $1.6$ atoms per spin component\,\cite{qu2020}.

\paragraph{Experimental results} 

\begin{figure}[t!]
	\centering
	\includegraphics[width=\columnwidth]{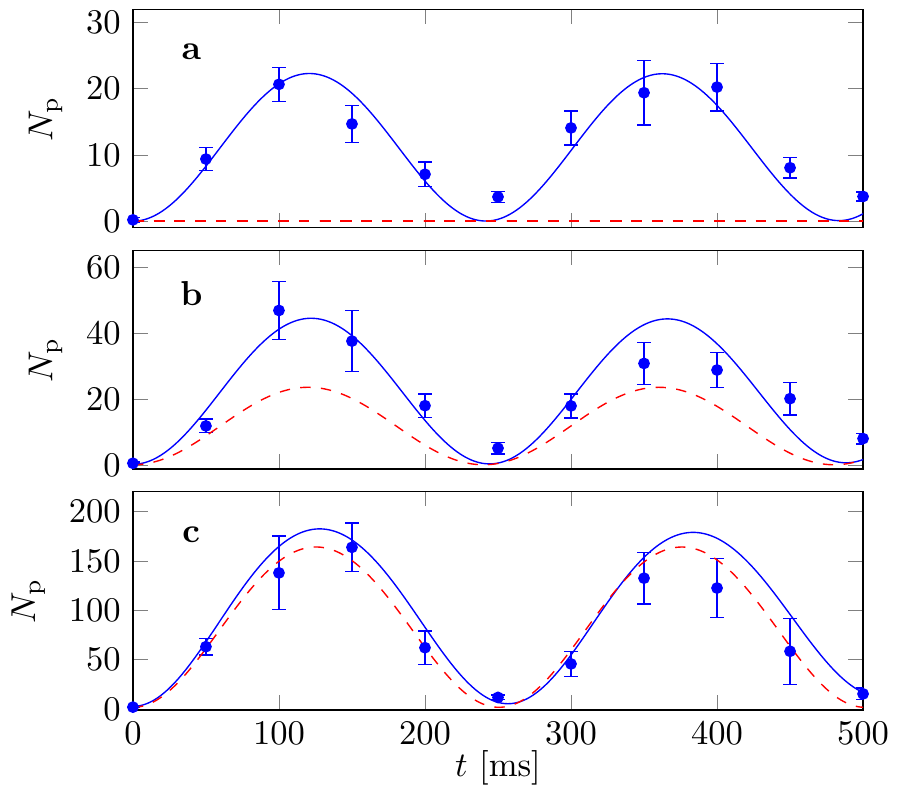}
	\caption{Evolution of the number of $(+1,-1)$ pairs $N_{\rm p}$ (circles) for $q/h\approx 0.22\pm0.03\,$Hz, $N\approx 1880\pm190$ atoms and various seed sizes: $N_{\rm seed}\approx0;\,0.25;\,1.8$ from (a) to (c). The initial phase is always set to $\theta_\ini\approx 0$. The solid lines are numerical solutions of the Schrödinger equation with the many-body Hamiltonian in Eq.(\ref{eq.Hamiltonian}) using $U_s/h = 9.9\,$Hz. The red dashed lines correspond to the classical prediction (\ref{eq: N_p_mf rev}). Here and in the following, error bars show the statistical error corresponding to two standard errors.}\label{fig1}
\end{figure}

In Fig.\,\ref{fig1}, we show the time evolution of $N_{\rm p}(t)$ for various initial states. In Fig.\,\ref{fig1}(a), we do not seed the dynamics. We observe an oscillation of $N_{\rm p}(t)$, not captured by the classical description of Eq.\,(\ref{eq: N_p_mf rev}), but in good agreement with the semi-classical predictions (\ref{eq: N_p_TWA rev}) or with the numerical resolution of the Schrödinger equation. In Fig.\,\ref{fig1}(b), we prepare a seed with $N_{\rm seed}\approx0.25\pm 0.03$ (inferred from a calibration of the rf power) and $\theta_i\approx0$. Compared to (a), the amplitude of the oscillations is doubled, in good agreement with (\ref{eq: N_p_TWA rev}). In Fig.\,\ref{fig1}(c), we set $N_{\rm seed}\approx 1.8\pm 0.2$ and $\theta_\ini\approx0$. The amplitude of the oscillations is further increased, and now also well reproduced by the fully classical treatment (\ref{eq: N_p_mf rev}).  In all cases (a,b,c), the condition $N_{\rm p}(t) \ll N$ remains fulfilled at all times. The validity of Eqs.\,(\ref{eq: N_p_mf rev},\ref{eq: N_p_TWA rev}) and the independence of the oscillation frequency on $N_{\rm seed}$ (as can be seen from Fig.\,\ref{fig1}) follow.

\begin{figure}[t!]
	\centering
	\includegraphics[]{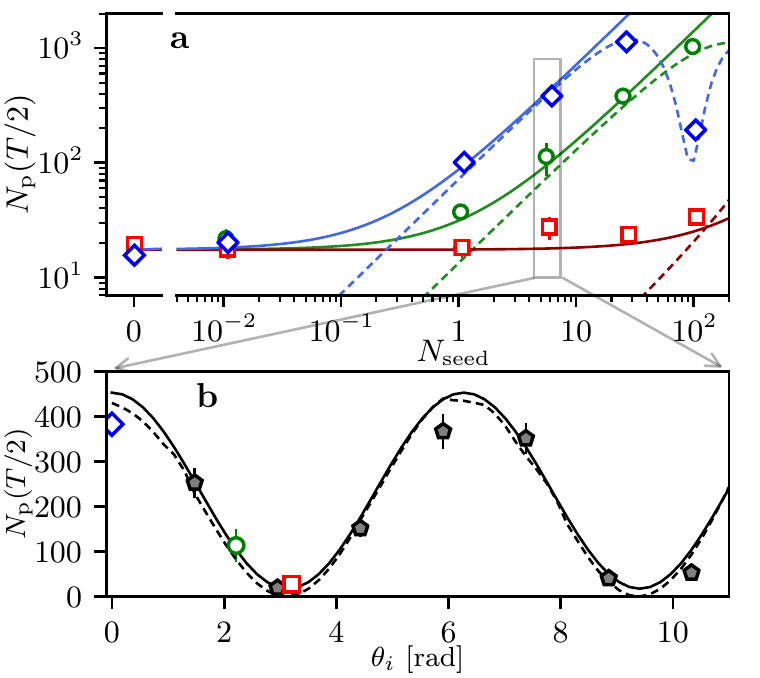}
	\caption{(a) Number of pairs produced after half a period of evolution versus $N_{\rm seed}$ for $q/h \approx 0.33\pm0.03\,$Hz and $N\approx2920\pm280$. The blue diamonds, green circles and red squares correspond to initial phases $\theta_\ini\approx 0$; $2.2$; and $3.3$\,rad, respectively. 
	For the three smallest seeds, $N_{\rm seed}$ is inferred from the calibration of the rf power. The solid lines are the semi-classical predictions given by Eq.\,(\ref{eq: N_p_TWA rev}) with $U_s/h \approx 12\,$Hz, assuming $N_{\rm p}\ll N$. For large $N_{\rm seed}$, this approximation breaks down, but a numerical solution of the non-linear classical mean-field Eqs.\,(\ref{eq: dnp/dt},\ref{eq: dtheta/dt}) with fixed initial conditions, becomes relevant. This fully classical treatment is shown as dashed lines.  (b) Scan of the initial phase $\theta_\ini$ after half a period of evolution for $N_{\rm seed}\approx 6.0\,$. }\label{fig2}
\end{figure}

We investigate the role of the initial phase $\theta_\ini$ in Fig.\,\ref{fig2}. In Fig.\,\ref{fig2}\,(a), we plot the variation of $N_{\rm p}(T/2)$, with $ T=\pi/\omega$ the period of oscillations, against $N_{\rm seed}$ for three values of $\theta_\ini$. For $N_{\rm seed}\ll 1$, we observe a saturation of $N_{\rm p}(T/2)$ at a value independent of $\theta_{\ini}$, consistent with the SC prediction (\ref{eq: N_p_TWA rev}). For such small seeds, the dynamics is triggered by quantum fluctuations. For larger seeds, unless the anti-phase-matching condition $\theta_\ini\approx\pi$ is fulfilled (red curves), stimulated emission becomes dominant and the fully classical description is accurate. We observe a linear increase of $N_{\rm p}(T/2)$ until the small-depletion approximation used to derive Eqs.\,(\ref{eq: N_p_mf rev},\ref{eq: N_p_TWA rev}) becomes inconsistent. For our data, this occurs for the point $N_{\rm seed}\approx100$\,, $\theta_\ini\approx0$. In this case, an exact resolution of the mean-field equations (\ref{eq: dnp/dt},\ref{eq: dtheta/dt}) provides accurate results. In Fig.\,\ref{fig2}\,(b), we set $N_{\rm seed}\approx6.0$ and scan the phase $\theta_\ini$. We measure oscillations of $N_{\rm p}(T/2)$ in good agreement with Eqs.\,(\ref{eq: N_p_mf rev},\ref{eq: N_p_TWA rev}).

\section{Relaxation dynamics}\label{sec: relaxation}

\begin{figure}
	\centering
	\includegraphics[width=\columnwidth]{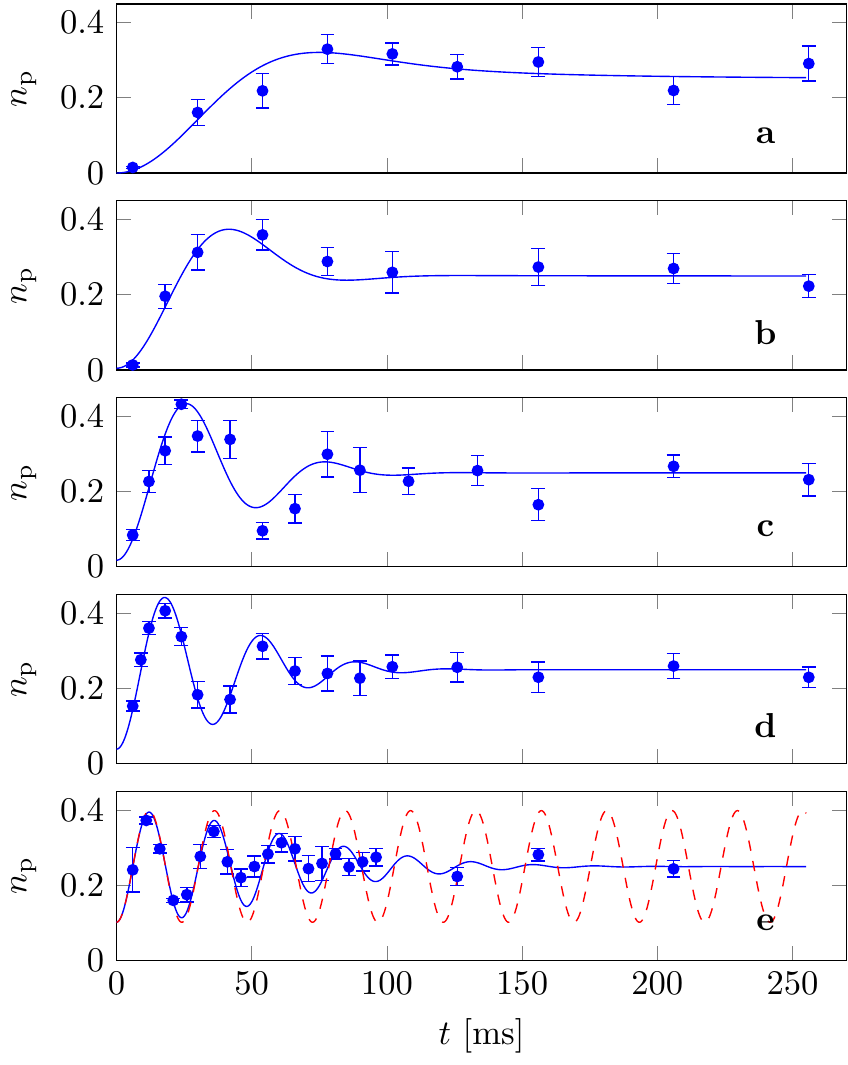}
	\caption{Evolution of the fraction of $(+1,-1)$ pairs $n_{\rm p}=N_{\rm p}/N$ in a negligible magnetic field, for $N\approx 124\pm12$ atoms and various seedings: $N_{\rm seed}=0;\,0.54;\,2.1;\,4.9;\,12.8;$ from (a) to (e). The initial phase is always set to $\theta_\ini\approx 0$. The solid lines are numerical solutions of the Schrödinger equation  for $U_s/h = 24.5\,$Hz. In (e), the red dashed line is the classical prediction from Eqs.\.(\ref{eq: dnp/dt},\ref{eq: dtheta/dt}).}\label{fig3}
\end{figure}

\paragraph{Theoretical prediction}
We now investigate the relaxation dynamics in a very small magnetic field, such that $q\ll U_s/N$. In this regime, the quadratic Zeeman shift $q$ is negligible and we set it to zero for the calculation. However, the assumption $n_{\rm p}\ll1$ used to derive Eq.\,(\ref{eq: N_p_mf rev}) is not valid and the mean-field equations\,(\ref{eq: dnp/dt},\ref{eq: dtheta/dt}) cannot be linearized. For $q=0$, the mean-field equations of motion can be solved directly. Taking for simplicity $\theta_\ini=0$, we find\,\cite{SM}
\begin{align}
	n_{\rm p}^{\rm (C)}(t)=\frac{1}{4}-\frac{1-4n_{\rm seed}}{4}\,\cos(\Omega t)\,,\label{eq: n_p_mf relax}
\end{align}
with an oscillation frequency
\begin{align}
	\Omega=\frac{4U_s}{\hbar} \,\sqrt{2n_{\rm seed}(1-2n_{\rm seed})}\,.\label{eq: Omega relax}
\end{align}

The non-linear dependence of $\Omega$ with $n_{\rm seed}$ reflects the non-linearity of the mean-field equations, and has dramatic consequences when one takes into account quantum fluctuations. The seeds spontaneously created from the vacuum of pairs induce random shifts of the oscillation frequency around its mean-field value. Averaging over many realizations therefore results in an intrinsic dephasing of the oscillations predicted in Eq.\,(\ref{eq: n_p_mf relax}). More precisely, for the generalized coherent spin state prepared in our experiment, the initial number of atoms in the $m=\pm1$ modes $N_{+1,\mathrm{ini}}+N_{-1,\mathrm{ini}}=\Sigma$ follows a binomial distribution of mean $2 N_{\rm seed}$ (quantum partition noise).
We use the random variable $\Sigma$ as an initial condition to solve the mean-field equations (\ref{eq: dnp/dt},\ref{eq: dtheta/dt}), \textit{i.e.} substituting $n_{\rm seed}$ in Eq.\,(\ref{eq: n_p_mf relax}) with $\Sigma/(2N)$. After averaging over the partition noise, 
we obtain for $N_{\rm seed}\gg 1$\,\cite{SM}
\begin{align}
	n_{\rm p}^{\rm (SC)}(t) \approx \frac{1}{4}-\frac{1-4n_{\rm seed}}{4}\cos(\Omega t)\ee^{-\frac{1}{2}(\gamma_{\rm c} t)^2}\,,\label{eq: n_p_avg relax}
\end{align}
with a collapse rate 
\begin{align}\label{eq:gammac}
\gamma_{\rm c}&= \frac{2U_s}{\sqrt{N}\hbar}\,\vert 1-4n_{\rm seed}\vert\,.
\end{align}
The analytic formula (\ref{eq: n_p_avg relax}) agrees very well with the numerical solution of the many-body Schrödinger equation for $N_{\rm seed}\gtrsim 1$.  The case $N_{\rm seed}\ll1$ can be treated using the truncated Wigner approximation \cite{mathew2017} or an exact diagonalization of the interaction Hamiltonian (\ref{eq.Hamiltonian_int}) \cite{evrard2020,law1998a}. The dynamics also displays a relaxation of $n_p$ to $1/4$, but with a different asymptotic behavior, $n_p-1/4\propto 1/t$. In a related work \cite{law1998b}, it was shown that Poissonian fluctuations of the atom number in each mode of a two-component BEC caused a Gaussian decay of the two-time correlation function. For the spin-1 and two-component cases, a similar mechanism is at work. The combination of non-linearities due to interactions and of quantum partition noise leads to dephasing and relaxation.

In an actual experiment, the relaxation of $N_{\rm p}$ is also enhanced by purely classical noise sources of technical origin. In our case, we identify shot-to-shot fluctuations of $U_s$ (see Section\,\ref{sec:spinorbec}) as a significant additional mechanism contributing to the blurring of the oscillations. To account for this phenomenon, we average Eq.\,(\ref{eq: n_p_avg relax}) over a Gaussian distribution of $U_s$ with variance $\delta U_s^2$. The resulting $n_{\rm p}(t)$ has the same functional form as in Eq.\,(\ref{eq: n_p_avg relax}) with the replacement 
\begin{align}
\gamma_{{\rm c}} \to \Gamma =\sqrt{\gamma_{\rm c}^2+\gamma_{\rm t}^2},\label{eq: gamma_c}
\end{align}
with a technical blurring rate 
\begin{align}
 \gamma_{\rm t}  = \frac{4\,\delta U_s}{\hbar} \, \sqrt{2n_{\rm seed}(1-2n_{\rm seed})}\,.\label{eq: Gamma_tot}
\end{align}
For small enough seeds $n_{\rm seed} \ll 1/4$, the total dephasing rate can be written
\begin{align}\label{eq:GammaSmallSeed}
 \Gamma & \approx \gamma_{\rm c} \sqrt{1+2\left(\frac{2\delta U_s}{U_s}\right)^2 N_{\rm seed}}.
\end{align}
This indicates a crossover from quantum to classical dephasing for seed sizes $N^\ast \approx  U_s^2/(2\delta U_s)^2$. 

\paragraph{Experimental considerations} 


In order to achieve the ``zero field'' regime $Nq \ll U_s$ experimentally, the best option is to reduce the atom number. Indeed, the density and therefore $U_s$, cannot be arbitrarily increased due to undesired inelastic processes. Reducing the applied magnetic field further is not feasible due to ambiant stray fields and environment-induced fluctuations (at the sub-mG level in our experiment).  Therefore, we lower $N$ by more than one order of magnitude with respect to the previous sections and prepare mesoscopic BECs of $N\approx 124\pm 12$ atoms. We also slightly tighten the trap in order to achieve $U_s/h \approx 24.5$\,Hz. In this case, the central spatial density remains low enough to avoid inelastic collisions  (more details in the Supplementary Material). 

\begin{figure}[t!]
	\centering
	\includegraphics[width=\columnwidth]{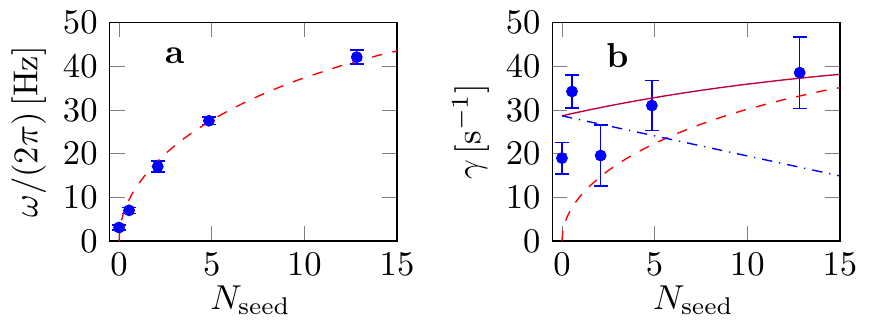}
	
	\caption{Frequency (a) and relaxation rate (b) of the spin-mixing dynamics in a negligible magnetic field. The circles are obtained from a fit to the data of Fig.\,\ref{fig3}, with the error bars indicating the $95\%$ confidence interval. In (a), the red dashed line corresponds to the frequency $\omega$ predicted by the mean-field treatment. In (b), the dash-dotted blue line corresponds to the rate $\gamma_{\rm c}$ of the collapse driven by quantum fluctuations, the red dashed line is the damping rate $\gamma_{\rm t}$ due to technical fluctuations,  and the solid purple line corresponds to the total damping rate $\Gamma=[\gamma_{{\rm c}}^2+\gamma_{\rm t}^2]^{1/2}$. We use the value $\delta U_s/U_s=0.13\pm 0.04$, obtained from a fit to the data.}\label{fig4}
\end{figure}

\paragraph{Experimental results} 

We show in figure\,\ref{fig3} the relaxation dynamics of $n_{\rm p}$ for various seed sizes $n_{\rm seed}$. We observe an acceleration of the initial dynamics for increasing $n_{\rm seed}$ and the emergence of rapidly damped oscillations. Eventually, $n_{\rm p}$ relaxes to the stationary value $\approx 1/4$ in all cases. Numerical simulations with $U_s$ taken as a fit parameter are overall in good agreement with the data, although they slightly underestimate the damping rate for the largest seed $N_{\rm seed}=12.8$.

To compare these experiments with the theoretical predictions, we fit a function of the form (\ref{eq: n_p_avg relax}) to the data of Fig.\,\ref{fig3}, leaving $\Omega$ and $\Gamma$ as free parameters. We report in Fig.\,\ref{fig4}a,b the fitted frequency and relaxation rate. The frequency is essentially insensitive to quantum or classical fluctuations, and the measured values agree well with the C or SC predictions. The relaxation rate varies little with $N_{\rm seed}$ in the range we have explored experimentally. This observation is explained by the SC theory including technical fluctuations. Indeed, the slow decrease of $\gamma_{\rm c}$ with $N_{\rm seed}$ is compensated by the increase of $\gamma_{\rm t}$. Using $\delta U_s/U_s \approx 0.13$ as determined in Fig.\,(\ref{fig4}), we find a ``quantum-classical crossover'' for seed sizes around $N^\ast \approx 15$, close to the largest value we explored experimentally. For small seeds $N_{\rm seed}\lesssim 5$, our measurements are consistent with a collapse driven primarily by quantum fluctuations. On the contrary, for the largest $N_{\rm seed}\approx 12.8$, classical technical dephasing is the dominant damping mechanism.

\section{Conclusion} 
We investigated the dynamics of a spin-1 BEC prepared with a majority of atoms in the Zeeman state $m=0$ and possibly small coherent seeds in the $m=\pm1$ modes. For a small but non-negligible magnetic field, we observe oscillations of the spin populations. This dynamics is triggered by quantum fluctuations in the absence of a seed, and cannot be captured in a completely classical approach. Adding a coherent seed is phase-sensitive \cite{wrubel2018}. In general it corresponds to a dramatic increase of the oscillation amplitude, and the classical predictions become accurate as soon as a few atoms (typically $N_{\rm seed} \gtrsim 2$) are used to seed the dynamics.

We also studied the dynamics in a negligible magnetic field. In this second regime, the combination of non-linear mean-field equations and quantum noise leads to the relaxation of the spin populations. When the size of the seed increases, the intrinsic damping rate $\gamma_{\rm c}$ decreases and the mean-field picture becomes more and more relevant. However, it eventually fails for sufficiently long times. Experimentally, technical noise sources provide additional dephasing mechanisms of purely classical origin that can be completely described in the mean-field approach. In our experiment, we identify the fluctuations of the total atom number as the leading blurring mechanism when the seed size exceeds a dozen atoms.

All the experiments presented in this Letter are well captured by a semi-classical theory, where quantum fluctuations are modeled using stochastic classical variables. An interesting direction for future work would be to test experimentally the validity of such a semi-classical description in other contexts, in particular in a chaotic regime \cite{evrard2020,Rautenberg2020,Tomkovic2017}.


\onecolumngrid
\newpage
\begin{center}
	\textbf{\large Supplemental Material: \\
		Coherent seeding of the dynamics of a spinor Bose-Einstein condensate:\\ from quantum to classical behavior }
	\vspace{0.cm}
\end{center}
\setcounter{equation}{0}
\setcounter{figure}{0}
\setcounter{table}{0}
\setcounter{page}{1}
\makeatletter
\renewcommand{\theequation}{S\arabic{equation}}
\renewcommand{\thefigure}{S\arabic{figure}}
\renewcommand{\bibnumfmt}[1]{[#1]}
\renewcommand{\citenumfont}[1]{#1}
\twocolumngrid

\setcounter{section}{0}
\section{Initial state preparation}

\subsection{Oscillating regime}

We prepare the spinor BEC at $t=0$ in a generalized coherent spin state $|\psi_\ini\rangle=\left(\sum_m\zeta_{\ini,m}|m\rangle\right)^{\otimes N}$, 
\begin{align}
{\bs \zeta}_\ini=\begin{pmatrix}
\sqrt{n_{\rm seed}}\ee^{i\frac{\theta_\ini+\eta_\ini}{2}}\\
\sqrt{1-2n_{\rm seed}}\\
\sqrt{n_{\rm seed}}\ee^{i\frac{\theta_\ini-\eta_\ini}{2}}\\
\end{pmatrix}\,.\label{SMeq: coherent stateSMSM}
\end{align}
We prepare this state starting from $\vert m=0 \rangle$ using a combination of magnetic field ramps and resonant radio-frequency (rf) pulses. In details, we first pulse a rf field resonant with the Zeeman splitting to populate the $m=\pm1$ modes with a fraction $n_{\rm seed}=\sin^2(\Omega_{\rm rf} t_1)/2$ of the atoms. Here, $\Omega_{\rm rf}$ is the rf Rabi frequency and $t_1$ the pulse duration. At this stage, we have prepared a coherent spin state of the form (\ref{SMeq: coherent stateSMSM}) with $\theta_\ini\approx\pi$. 

To change $\theta_\ini$, we let the system evolve in a field $B=0.5\,$G ($q/h\approx70\,$Hz) for a time $t_2<h/(2q)$, before quenching the magnetic field down to $28\pm2\,$mG ($q/h\approx0.22\,$Hz) in $t_3=4\,$ms to achieve the desired regime $U_s/N\ll q\ll U_s$. Interactions are negligible ($U_s/h\approx 10\,$Hz hence $U_st_{2,3}/h\ll1$), and the system simply acquires a phase shift $\Delta\theta_2=-2qt_2/\hbar$ while the magnetic field is held constant, and $\Delta \theta_3=-2\int q(t)dt/\hbar$ during the quench. This results in an initial phase $\theta_{\rm ini}=\pi-2qt_2/\hbar+\Delta\theta_3$ that is fully tunable from 0 to $2\pi$ by varying $t_2$.

\subsection{Relaxing regime}

We prepare mesoscopic BECs of $N\approx124$ atoms in the same initial spin state as before. We lower the magnetic field down to $B=4.2\pm1.5$\,mG ($q/h\approx5\,$mHz) in $t_3=20\,$ms. The ramp time corresponds to the time needed for the damping of eddy currents in the vacuum chamber. Because of the small atom number, the effects of the spin dependent interactions are negligible over the ramp ($U_s/h\approx4\,$Hz, such that $U_st_3/h\ll1$) and the evolution of the state is essentially another phase shift of $\theta$, which can be compensated for by varying $t_2$.
For these experiments, we always choose $t_2$ such that $\theta_\ini\approx0$. 

Finally, we trigger the dynamics by recompressing the trap in $6\,$ms ($U_s/h\approx 4\to24\,$Hz). By performing numerical simulations of the sequence with the many-body Schr\"odinger equation, we have checked that the ramp can be considered instantaneous to a good approximation. 

\section{Classical and semi-classical dynamics}

We detail here the calculations of the dynamics of $N_{\rm p}(t)$ given in the main text. We use a classical (C) approach based on the mean-field approximation and a semi-classical (SC) approach inspired by the truncated Wigner approximation (TWA). In both frameworks, the annihilation operators $\hat a_{ m}$ are replaced by $c$-numbers $\alpha_m=\sqrt{N}\zeta_m$, with $N$ the number of condensed atoms and $\bm{\zeta}$ a spin-1 wavefunction (normalized to unity) parameterized as
\begin{align}
{\bs \zeta}=\begin{pmatrix}
\sqrt{n_{\rm p}}\ee^{i\frac{\theta+\eta}{2}}\\
\sqrt{1-2n_{\rm p}}\\
\sqrt{n_{\rm p}}\ee^{i\frac{\theta-\eta}{2}}\\
\end{pmatrix}\,.\label{SMeq: coherent stateSM}
\end{align}
Here $n_{\rm p}=(N_{+1}+N_{-1})/(2N)$ denotes the average number of $m=\pm 1$ pair normalized to the total atom number ($N_{\rm p}=N n_{\rm p}$), and we have restricted ourselves to the situation $N_{+1}=N_{-1}$. We also have chosen $\zeta_0$ real without loss of generality. 

The mean field equations of motion for a spin-1 condensate in the single-mode regime take the form\,\cite{zhang2005aSM,kawaguchi2012SM}
\begin{align}
\hbar\dot n_{\rm p}&=-2U_sn_{\rm p}(1-2n_{\rm p})\sin\theta\,\label{SMeq: dnp/dt}\\
\hbar\dot \theta&=-2q+2U_s(4n_{\rm p}-1)(1+\cos\theta)\,.\label{SMeq: dtheta/dt}
\end{align}
The mean-field energy per atom is given by 
\begin{align}
\mathcal{E}_s=2U_sn_{\rm p}(1-2n_{\rm p})(1+\cos\theta)+2qn_{\rm p}\,.\label{SMeq: Emf}
\end{align}
The energy $\mathcal{E}_s$ is a constant of motion, a fact that we will used repeatedly in the following.

\subsection{Dynamics in the oscillating regime}\label{sec. rev regime}
In this section we derive the evolution of $N_{\rm p}(t)$ for the oscillating regime $q\gg U_s/N$. We assume $N_{\rm seed}\ll N$, \textit{i.e.} the situation where quantum fluctuations may play a significant role. For $N_{\rm seed}\sim N$, a fully classical treatment is accurate.

\paragraph{Classical solution :}
Assuming $n_{\rm p}\ll 1$, we linearize Eqs.\,(\ref{SMeq: dnp/dt}) and (\ref{SMeq: Emf}), 
\begin{align}
\hbar\dot n_{\rm p}& \approx -2U_s n_{\rm p}\sin\theta\,\\
\mathcal{E}_s&  \approx  \Big(2U_s (1+\cos\theta)+2q \Big)n_{\rm p}\,.
\end{align}
We use the second equation to express $\cos\theta$ as a function of $n_{\rm p}$ and of the constants $q,U_s,\mathcal{E}_s$. Substituting in the first equation, we obtain a differential equation on $n_{\rm p}$ only, $\dot n_{\rm p}^2=-4\omega^2\left[n_{\rm p}-\alpha\right]^2+A$, where 
\begin{align}
\hbar\omega=\sqrt{q(q+2U_s)}, \hspace{0.5cm} \alpha=\frac{\mathcal{E}_s(q+U_s)}{2(\hbar\omega)^2},
\end{align} 
and where $A$ is constant. Differentiating one more time, we find that either $n_{\rm p}$ is constant or it obeys the harmonic equation $\ddot n_{\rm p}+4\omega^2\left(n_{\rm p}-\alpha\right)=0$.
The evolution is thus a harmonic motion at frequency $2\omega$, 
\begin{align}
n_{\rm p}(t) \approx n_{\rm seed} +2(\alpha-n_{\rm seed})\sin^2(\omega t) \,,\label{SMeq: dnp rev}
\end{align}
with the initial conditions $n_{\rm p}(0)=n_{\rm seed}$ and  $\theta(0)=\theta_\ini$. 

If we also assume (as in the experiments we performed) that $q\ll U_s$, we have $\mathcal{E}_s\approx 4U_s n_{\rm seed}\cos^2(\theta_\ini/2) \gg q$, and $\alpha \approx \mathcal{E}_s/(4q) \gg 1$. Eq.\,(\ref{SMeq: dnp rev}) then reduces to 
\begin{align}
n_{\rm p}(t) \approx n_{\rm seed} + \frac{2 U_s n_{\rm seed}}{q} \, \cos^2(\theta_\ini/2)\,\sin^2(\omega t) \,,\nonumber
\end{align}
\textit{i.e.} to Eq.\,(\ref{eq: N_p_mf rev}) in the main text.


\paragraph{Semi-classical picture :}

We now consider the effect of quantum fluctuations within the TWA\, \cite{polkovnikov2010SM,steel1998SM,sinatra2002SM,mathew2017SM,wrubel2018SM}. In this method, the $c$-numbers $\alpha_m$ used instead of the annihilation operators $\hat a_m$ in the mean-field approximation are treated as complex random variables. At $t=0$, these variables sample the Wigner distribution of the initial state $\vert\psi_\ii\rangle$. Their mean values are given by
\begin{align}
\bar{\bs \alpha}_\ini=N\begin{pmatrix}
\sqrt{n_{\rm seed}}\,\ee^{i\frac{\theta_\ini+\eta_\ini}{2}}\\
\sqrt{1-2n_{\rm seed}}\\
\sqrt{n_{\rm seed}}\,\ee^{i\frac{\theta_\ini-\eta_\ini}{2}}\\
\end{pmatrix}\,.
\end{align}
In the limit $N_{\rm seed}\ll N$, the calculation can be simplified by neglecting the depletion of the mode $m=0$. For the $m=\pm1$ modes, this approximation amounts to replacing coherent spin states by harmonic oscillator coherent states, which are considerably easier to handle. The initial quantum state is thus taken to be
\begin{align}
\vert\psi_\ini\rangle  \approx \frac{1}{\sqrt{N!}}\prod_{m=\pm1}\ee^{\bar \alpha_{m,\ini}\hat a_m^\dagger-\bar \alpha_{m,\ini}^*\hat a_m}\hat{a}_0^{\dagger N}|{\rm vac}\rangle\,.
\end{align}

For $t>0$, the equations of evolution (\ref{SMeq: dnp/dt},\ref{SMeq: dtheta/dt}) remain valid in the TWA. The solution for initial conditions $\alpha_{\pm1,\ini}$ is thus given by Eq.\,(\ref{SMeq: dnp rev}) with the substitution $4N_{\rm seed}\cos^2(\theta_\ini/2) \to \vert\alpha_{+1,\ini}+\alpha_{-1,\ini}^{*}\vert^2$. 

To average over the initial distribution of $\alpha_{\pm1,\ini}$, we recall that the Wigner distribution average $\mean{\mathcal{O}(\alpha_m,\alpha_m^*)}_{\rm Wig}$ of an operator $\mathcal{O}$ is equal to the expectation value $\mean{\mathcal{O}^{\rm sym}(\hat a_m,\hat a_m^\dagger)}$  of the corresponding symmetrically ordered operator $\mathcal{O}^{\rm sym}$ \cite{polkovnikov2010SM}. We obtain
\begin{align}
&\mean{\alpha_{+1,\ini}\alpha_{-1,\ini}^{\ast}}_{\rm Wig}=\mean{\hat{a}_{+1}\hat{a}_{-1}^\dagger}=\bar\alpha_{+1,\ini}\bar\alpha_{-1,\ini}^\ast\,,\\
&\mean{\vert\alpha_{m,\ini}\vert^2}_{\rm Wig}=\frac{1}{2}\mean{\hat a^\dagger_m\hat a_m+\hat a_m\hat a_m^{\dagger}}=\vert \bar\alpha_{m,\ini}\vert^2+\frac{1}{2}\,.
\end{align}
This leads to 
\begin{align}
\langle N_{\rm p}(t)\rangle \approx \frac{U_s}{2q}\sin^2(\omega t) \,\left(\vert \bar\alpha_{+1,\ini}+\vert \bar\alpha_{-1,\ini}^\ast\vert^2+1 \right)\,,\nonumber
\end{align}
which gives Eq.\,(\ref{eq: N_p_TWA rev}) in the main text.

As a final remark, we note that the Bogoliubov method is also well suited to study the regime that we investigated here, and leads to the same result \cite{mias2008SM,uchino2010aSM,evrard2020SM}.

\subsection{Relaxation dynamics}

We now discuss the regime $q\ll U_s/N$, in which we observe a relaxation of the number of pairs $N_{\rm p}$ to a stationary value. In this regime, the quantum fluctuations play an important role even for $N_{\rm seed}\gg 1$. We will thus consider that $N_{\rm seed}\gg 1$ and $N-N_{\rm seed}\gg 1$. For simplicity, we will focus on the situation $\theta_\ini=0$, for which the effect of the seed is maximal. The case with no seed has been treated using an exact diagonalization of the Hamiltonian\,\cite{evrard2020SM} or the TWA\,\cite{mathew2017SM}.

\paragraph{Classical solution} 

In order to simplify the calculation, we neglect completely the quadratic Zeeman shift. In this regime $q\ll U_s/N$, the Zeeman term indeed plays no significant role even for the fully quantum model. Introducing the auxiliary variable $x=4n_{\rm p}-1$, the equations of motion and the energy become
\begin{align}
\hbar \dot x&=-U_s(1-x^2)\sin\theta\,,\\
\hbar\dot \theta&= 2U_s x (1+\cos\theta)\,,\\
\mathcal{E}_s&=\frac{U_s}{4}(1-x^2)(1+\cos\theta)=\mathrm{cst}\,.
\end{align}
We combine the first and last equations to obtain
\begin{align}
\dot x&=-\frac{4\mathcal{E}_s}{\hbar}\frac{\sin\theta}{1+\cos\theta}.
\end{align}
Differentiating this equation, we eliminate the phase $\theta$ and obtain a simple harmonic equation, $\ddot x=-\Omega^2 x$, with an oscillation frequency $\hbar\Omega=\sqrt{8U_s\mathcal{E}_s}$. For the initial conditions $n_{\rm p}(0)=n_{\rm seed}$ and $\theta(0)=0$, we have $\hbar\Omega=2U_s\sqrt{1-x_0^2}$ and $x(t)=x_0\cos(\Omega t)$ with $x_0=4n_{\rm seed}-1$. This corresponds to the results announce in Eqs.\,(8,9) of the main text.

\paragraph{Quantum partition noise:}
The initial state
\begin{align}
|\psi_\ini\rangle=\frac{1}{\sqrt{N!}}\left[\sum_{m=0,\pm 1}\zeta_m\,\hat{a}^{\dagger}_{m}\right]^N|{\rm vac}\rangle\,,\nonumber
\end{align}
is characterized by fluctuations of the number of $\pm 1$ atoms. We consider again the states with $\vert \zeta_{+1} \vert=\vert \zeta_{-1} \vert=\sqrt{N_{\rm seed}}$ and $\theta_\ii=0$. We introduce the sum $\Sigma=N_{+1}+N_{-1}$, its relative value $s=\Sigma/N$ and the difference $\Delta=N_{+1}-N_{-1}$. The components of $\bm{\zeta}$ are related to the average $\bar{\Sigma}$ of $\Sigma$ by
\begin{align}
\vert \zeta_{\pm 1} \vert^2=\frac{\bar{\Sigma} }{2}, \hspace{0.5cm}
\vert \zeta_{0} \vert^2=N-\bar{\Sigma}.
\end{align}
The joint distribution of $\Sigma$ and $\Delta$ in the initial coherent spin state is 
\begin{align}\label{SMeq:PSD}
\mathcal{P}(\Sigma,\Delta)&=\frac{N!}{\left(\frac{\Sigma+\Delta}{2}\right)! \left(\frac{\Sigma-\Delta}{2}\right)! (N-\Sigma) !} \left(\frac{\bar{s}}{2}\right)^\Sigma (1-\bar{s})^{N-\Sigma}.
\end{align}
We deduce from Eq.\,(\ref{SMeq:PSD}) the distribution of $\Sigma$,
\begin{align}
\mathcal{P}(\Sigma)&=\frac{N!}{\Sigma ! (N-\Sigma) !} \bar{s}^\Sigma (1-\bar{s})^{N-\Sigma}.
\end{align}
with $\Sigma \in  [0,N]$. The normalization follows from the binomial formula.

For large $N$ and $\Sigma$ away from the extreme values $0, N$, the binomial distribution is well approximated by a continuous Gaussian distribution 
\begin{align}\label{SMeq: P(x)}
\mathcal{P}(\Sigma)&\approx \frac{1}{N}\frac{1}{\sqrt{2\pi}\sigma} \ee^{  
	-\frac{(s-\bar{s})^2}{2\sigma^2}}=p(s) ds.
\end{align}
with a step size $ds=1/N$ and a standard deviation
\begin{align}
\sigma & = \sqrt{\frac{\bar{s}(1-\bar{s})}{N}}=\sqrt{\frac{2n_{\rm seed}(1-2n_{\rm seed})}{N}}.
\end{align}
One can check the normalization of both distributions,
\begin{align*}
\sum_{\Sigma=0}^{N} \mathcal{P}(\Sigma)& \to \int_0^{1} f(s) ds \approx \int_{-\infty}^{+\infty} \frac{1}{\sqrt{2\pi}} \ee^{  
	-\frac{u^2}{2}} du=1.
\end{align*}
To extend the lower boundary to $-\infty$, we require $\bar{s}/\sigma=\sqrt{N}\times \sqrt{\bar{s}/(1-\bar{s})} \gg 1$, or $N \bar{s} =2N_{\rm seed} \gg 1$.
%

\paragraph{Semi-classical picture of the dynamics:}

Similarly to what we have done in Sec.\,\ref{sec. rev regime}, we average the mean field solution (\ref{SMeq: dnp/dt},\ref{SMeq: dtheta/dt}) with $2n_{\rm seed}\to s$ over the probability distribution $p(s)$ in Eq.\,(\ref{SMeq: P(x)}). This amounts to compute the integral
\begin{align}
I=\frac{1}{2} \int_{0}^{1} \,s\,  \cos[\Omega(s)t]\, p(s)\,ds.\label{SMeq: I}
\end{align}
We use the fact that $p(s)$ is sharply peaked around $\bar{s}$, with a width $\sim 1/N$ much narrower than the scale of variation of the rest of the integrand $s\cos[\Omega(s)t]$. As a result, we extend the integral boundaries to $\pm\infty$, set $s \approx \bar{s}$ and expand the frequency $\Omega(s)$ to first order,
\begin{align}
\Omega(s) \approx \bar{\Omega}+\bar{\Omega}' (s-\bar{s}) +\mathcal{O}(\epsilon^2)\,,
\end{align}
where $\bar\Omega=\Omega(\bar{s})$ and $\bar\Omega'=\Omega'(\bar{s})=(2 U_s /\hbar)\times (1-2\bar{s})/\sqrt{\bar{s}(1-\bar{s})}$.

With straightforward manipulations, we cast $I$ in the form of the Fourier transform of a Gaussian function, which is readily calculated. We find
\begin{align}
I & =\frac{1}{2}\bar{s}\cos[\bar{\Omega}t] \, \ee^{-\frac{1}{2}(\gamma_{\rm c} t)^2},
\end{align}
with a damping rate
\begin{align}
\gamma_{\rm c} =\vert \bar{\Omega}'\sigma\vert=\, \frac{2U_s}{\sqrt{N}\hbar} \, \vert 1-2\bar{s}\vert .
\end{align}
Using $\bar{s}=2n_{\rm seed}$, this gives Eq.\,(11) in the main text.

\paragraph{Classical fluctuations of $\Omega$:}
In addition to the intrinsic dephasing originating from quantum fluctuations, any technical fluctuations of $\Omega$ will also contribute to the observed relaxation. We consider here the dominant source of classical blurring in our experiment, namely fluctuations of the interaction strength $U_s$ mainly due to shot-to-shot atom number fluctuations.

We model these fluctuations by considering a fluctuating interaction strength $U_s'=U_s +\delta U_s x$, with $U_s$ the average value, $ \delta U_s$ the standard deviation of the noise, and $x$ a centered Gaussian random variable of variance unity. This leads to a fluctuating oscillation frequency $\Omega(x)=\bar{\Omega}(1+x \cdot \delta U_s/U_s)$. We neglect the fluctuations of $\gamma_{\rm c}$, which is legitimate for $N_{\rm seed}\gg1$ and hence $\gamma_{\rm c}\ll\bar\Omega$.
Averaging over the Gaussian probability distribution $p(x)$, we find that
\begin{align}
I_2 & =\left\langle\,\cos[\Omega(x)t]\ee^{-\frac{1}{2}(\gamma_{\rm c}t)^2}\, \right\rangle_{x}=\cos[\bar{\Omega}t]e^{-\frac{1}{2}(\gamma_{\rm t}t)^2-\frac{1}{2}(\gamma_{\rm c}t)^2},\label{SMeq: I2}
\end{align}
with a classical (technical) damping rate given by
\begin{align}
\gamma_{\rm t}&  = \bar{\Omega} \frac{\delta U_s}{U_s}.\label{SMeq: gamma_exp}
\end{align}
From Eqs.\,(\ref{SMeq: I2},\ref{SMeq: gamma_exp}) we obtain Eqs.\,(\ref{eq: gamma_c},\ref{eq: Gamma_tot}) given in the main text.

\end{document}